\documentclass[conference]{IEEEtran}

\usepackage{graphicx}
\graphicspath{{./}{figures/}}
\usepackage{pgfplots}
\pgfplotsset{compat=newest}
\usepackage[utf8]{inputenc}
\usepackage{amsmath}
\usepackage{refcount}
\usepackage{cite}

\usepackage{xcolor}

\newcommand{\Figure}[1]{Figure~\ref{#1}}
\newcommand{\Section}[1]{Section~\ref{#1}}
\usepackage{microtype}

\usepackage{booktabs}
\usepackage{color}
\usepackage{url}
\usepackage{makecell} 
\usepackage{enumitem}
\usepackage{float}
\usepackage{soul}
\usepackage{dblfloatfix} 

\usepackage[pscoord]{eso-pic}
\newcommand{\placetextbox}[3]{
\setbox0=\hbox{#3}
\AddToShipoutPictureFG*{
\put(\LenToUnit{#1\paperwidth},\LenToUnit{#2\paperheight}){\vtop{{\null}\makebox[0pt][c]{#3}}}%
}%
}%

\title{Exploring Perceptual Audio Quality Measurement on Stereo Processing Using the Open Dataset of Audio Quality}

\author{
\IEEEauthorblockN{
Pablo M. Delgado\IEEEauthorrefmark{1},
Sascha Dick\IEEEauthorrefmark{1},
Christoph Thompson\IEEEauthorrefmark{2},
Chih-Wei Wu\IEEEauthorrefmark{3},
Phillip A. Williams\IEEEauthorrefmark{3}}
\IEEEauthorblockA{\IEEEauthorrefmark{1}Fraunhofer Institute for Integrated Circuits IIS, Erlangen, Germany}
\IEEEauthorblockA{\IEEEauthorrefmark{2}Ball State University, Muncie, USA}
\IEEEauthorblockA{\IEEEauthorrefmark{3}Netflix, Inc., Los Gatos, USA}
}

\begin{document}

\maketitle

\begin{abstract}
ODAQ (Open Dataset of Audio Quality) provides a comprehensive framework for exploring both monaural and binaural audio quality degradations across a range of distortion classes and signals, accompanied by subjective quality ratings. A recent update of ODAQ, focusing on the impact of stereo processing methods such as Mid/Side (MS) and Left/Right (LR), provides test signals and subjective ratings for the in-depth investigation of state-of-the-art objective audio quality metrics. Our evaluation results suggest that, while timbre-focused metrics often yield robust results under simpler conditions, their prediction performance tends to suffer under the conditions with a more complex presentation context. Our findings underscore the importance of modeling the interplay of bottom-up psychoacoustic processes and top-down contextual factors, guiding future research toward models that more effectively integrate both timbral and spatial dimensions of perceived audio quality.
\end{abstract}

\placetextbox{0.5}{0.08}{\fbox{\parbox{\dimexpr\textwidth-2\fboxsep-2\fboxrule\relax}{\footnotesize \centering Accepted for presentation at the Audio Engineering Society (AES) 159th Convention, October 2025, Long Beach, USA.}}}

\section{Introduction}
The Open Dataset of Audio Quality (ODAQ) \cite{Torcoli2024ODAQ} is an evolving repository designed to examine both monaural and binaural aspects of audio quality degradation through a diverse set of signals and distortion types, and their accompanying subjective quality ratings. Currently, the collection features audio samples processed by multiple distortion classes in stereo format. The dataset is meant to provide fundamental material for exploring how distinct audio artifacts affect perceived audio quality.

\begin{figure*}[t]
\centering
\resizebox{\textwidth}{!}{%
\includegraphics{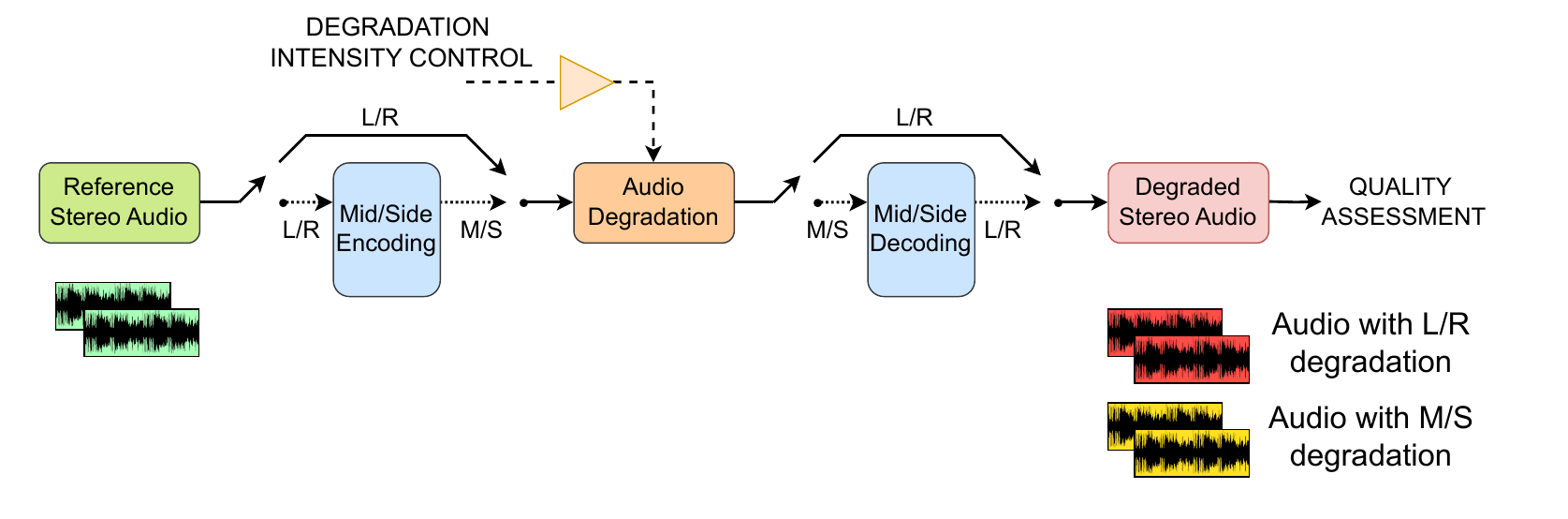}
}
\caption{Experimental setup for the generation and quality assessment of the degraded audio files.}
\label{fig:SignalGeneration}
\end{figure*}

A primary motivation for ODAQ is its potential to act as a benchmark for existing and novel objective audio quality metrics. Although ODAQ’s initial performance on several timbre-based quality metrics has already been briefly reported \cite{dick2024ODAQ}, it would be valuable to extend its capabilities to include binaural perceptual models and their corresponding timbre and spatial audio quality metrics. In this context, this paper serves as an example analysis of the latest update of the ODAQ database to address this need. In its latest update, ODAQ supplies test signals and subjective ratings that feature degradations introduced using two stereo processing techniques (see \Figure{fig:SignalGeneration}). By utilizing this new material, this work aims to inspire further research that makes use of ODAQ's comprehensive framework for exploring both monaural and binaural audio quality degradations

Moreover, the necessity of such research comes into focus when considering that the scope of spatial audio quality modeling has remained limited to only certain aspects of quality degradation, such as changes in perceived localization of auditory objects, envelopment, width sensation and amount of externalization \cite{BlauertBook}. While prior work has concentrated on studying these aspects separately, or in combination with timbre aspects \cite{rumsey2005relative, MoBiQ}, several critical issues remain underexplored. For example, little is known about how audio quality metrics perform when the same distortion signal appears in both channels simultaneously, or when independent, channel-specific distortions arise as the result of established audio coding and processing techniques such as Mid/Side (MS) or Left/Right (LR) channel coding, respectively. These distinct modes of operation can result in different perceptual outcomes, particularly regarding stereo imaging and artifact audibility, making them important scenarios for investigation.

Another underexamined element is the role of presentation context in shaping subjective quality judgments \cite{zielinski:2008, MUSHRA, PEAQ, PEMOQ}. Typically, perceptual models focus on bottom-up psychoacoustic processes such as critical band analysis, auditory masking and modulation strength for measuring artifact detectability and perceived signal degradation \cite{PEAQ, PEMOQ, delgado2024towards,HAAQI}, and inter-aural signal differences for estimating distortions in the evoked spatial image \cite{MoBiQ,eMoBiQ, delgado2023design, seo2013perceptual,kmpf2010standardization}. Yet, in real listening tests, higher-level factors—such as the test format, the anchor conditions, and whether multiple quality levels are presented concurrently—can dramatically influence listeners’ judgments \cite{zielinski:2008, MUSHRA}. Based on the unique design paradigm in ODAQ, which allows for the presentation of the same set of distortions in varied contexts, this study focuses on different stereo processing approaches and presentation modes. In this scenario, we investigate if the interaction of top-down and bottom-up perceptual processes as observed in human ratings is also reflected on the studied objective audio quality metrics.

\section{Method}

We review several established audio quality metrics that have been extensively validated in various tasks beyond assessing the quality of perceptually coded audio signals, such as psychoacoustic experiments \cite{PEMOQ} and speech intelligibility \cite{HAAQI}. These metrics also include documented binaural extensions \cite{delgado2023design, MoBiQ} for evaluating spatial audio quality, in addition to timbre quality. To validate these systems, we assess their quality prediction performance by comparing their outputs with the subjective quality scores from the database presented in \cite{dick2025ODAQ}, with a particular emphasis on linearity \cite{EvalObjective}.

\subsection{Studied metrics}
\label{sec:metrics}

The metrics in this study are produced by intrusive perceptual objective audio quality assessment systems that calculate internal representations of a reference and a possibly degraded digital audio signal, the signal under test (SUT). The internal representations are informed on psychoacoustic and physiological models of human hearing. Based on these representations, the metrics compute perceptually-motivated distances that have been shown to correlate more strongly with perceived quality degradation intensities and disturbance detection thresholds than traditional signal-based quality metrics such as Signal-to-Noise Ratio (SNR) \cite{brandenburg1987evaluation}. The studied quality assessment systems are widely used in applications that range from perceptual audio coding, speech-intelligibility assessment to hearing-aid audio quality evaluation and basic psychoacoustic research.

The evaluated metrics are:

\begin{itemize}
    \item Perceptual Evaluation of Audio Quality (PEAQ) in its \textit{basic} operational mode, we analyze the Objective Difference Grade (ODG) output~\cite{PEAQ}\footnotemark[\getrefnumber{noteKabal}].
    \item The Noise-to-Mask Ratio (NMR) metric~\cite{brandenburg1987evaluation}\footnote{\label{noteKabal} As implemented in \cite{Kabal}, available at: \url{https://www.mmsp.ece.mcgill.ca/Documents/Software/}}
    \item The \mbox{2f-model}~\cite{ksrWaspaa19}, based on the same PEAQ implementation.
    \item The Hearing-Aid Audio Quality Index (HAAQI), operating in Normal Hearing (NH) mode. \cite{HAAQI} \footnote{ Implementation of HAAQI v1, available at: \url{https://github.com/claritychallenge/clarity}}
    \item The PEMO-Q quality assessment system's ODG (low-pass, `lp' and filterbank `fb' versions) \cite{PEMOQ} \footnote{\mbox{ }Implementation available at: \url{https://uol.de/mediphysik/downloads/pemo-q}}
    \item The MoBi-Q model, with a monaural core for timbre quality estimation (GPSMq) and binaural extension (binQ) for spatial quality estimation \cite{MoBiQ},
    \item The eMoBi-Q model, featuring an efficient model for monaural and binaural estimation of audio quality based on MoBi-Q \cite{eMoBiQ},
    \item PEAQ-CSM (Cognitive Salience Model), an extension of PEAQ using a cognitive model for distortion salience  \cite{delgado2024towards},
    \item Two binaural extensions of PEAQ based on DFT- and Filterbank-based analysis binaural cue extraction models \cite{delgado2023design}.
\end{itemize} 

All of the analyzed metrics can process at least two-channel audio, but only a few include an explicit binaural model to evaluate spatial-image distortions—most handle each channel independently.

In the following, we omit a detailed overview of the perceptual models behind each objective metric as these are well covered in the cited literature. Instead, the next sections focus on the relevant aspects of their binaural processing for spatial audio quality prediction, and how this processing interacts with the respective timbre-quality assessment modules.

\subsection{Handling of stereo signals}
\label{sec:HandlingOfStereoSignals}
All methods report a quality estimate based on the values of one or more distortion metrics averaged over time and frequency with possible perceptual weightings, or max/min values, over time and frequency. In the case of HAAQI, the reported quality score is the averaged quality index of the left and the right channel over the total duration of the signal. 

In the case of PEAQ, the multiple distortion metrics (called Model Output Variables - MOVs) are linearly averaged between the left and right channel after the temporal averaging. For producing PEAQ's ODG scores, the channel averages are then fed into an Artificial Neural Network (ANN) to map them to a final quality (ODG) score between 0 (no impairment of SUT) and -4 (strong impairment). Similarly, the 2f and PEAQ-CSM variants map MOVs to a unique quality score based on the channel averages of the individual MOVs. 

The only exception for a separate treatment of audio channels in the system is given by the calculation of loudness and energy minimum thresholds for the calculation of certain MOVs (the minimum of both channels), or for the establishment of data boundaries in the pre-processing step (to remove a noise floor in the SUT present before the start and end of the comparison to the reference). 

In MoBi-Q, the monaural GPSMq (Generalized Power Spectrum Model for quality) feature in charge of estimating timbre quality also processes the left and right channels of a binaural signal independently and then averages their outputs. However, GPSMq was modified to be less sensitive to binaural distortions caused by long-term inter-channel level differences: a processing front-end normalizes these differences before entering the model so that the monaural model estimator effectively becomes insensitive to inter-aural level difference (ILD) cues. In the case of eMoBi-Q, left and right channels are concatenated. The resulting timbre quality metric based on GPSMq that handles stereo files with spatial distortion was dubbed OPMdual.

For PEMO-Q, we average the ODQ values across channels for the quality estimate, as it has been shown that ODG outperforms other metrics based on the same model \cite{harlander2014sound}. 

\subsubsection{MoBI-Q: monaural and binaural models}

MoBi-Q predicts overall audio quality by combining monaural and binaural cues within two perceptual-model frameworks. On the monaural side for estimation of timbre quality, it uses a front end (GPSMq) consisting of outer/middle-ear filtering, a gammatone filterbank, cochlear compression, half-wave rectification, and low-pass hair-cell filtering, and then splits each band into fine-structure ($<$ 1.4 kHz) and envelope ($>$ 1.4 kHz) paths to compute spectral-coloration and loudness metrics. In parallel, its binaural branch adopts a binaural  psychoacoustic front end (BAM-Q) on the same filter outputs to derive ILDs, interaural time/phase differences (ITD/IPD), and interaural coherence expressed through inter-aural vector strengths (IVS). All sub-measures are calculated in consecutive time frames, averaged across bands and time.

\subsubsection{eMoBI-Q: monaural and binaural models}

The eMoBi-Q system is an efficient monaural and binaural audio quality metric based on MoBi-Q that is thought for time-critical applications such as real time control of algorithms in audio and hearing technology. It replaces MoBi-Q's computationally intensive binaural model (BAM-Q) with only two binaural cues: the complex correlation coefficient $\gamma$ (capturing coherence and inter-aural phase difference) and inter-aural level differences. It  extracts both monaural and binaural features from a single peripheral filterbank, and retains only the linear component of GPSMq for spectral coloration and loudness. Additionally, the model unifies the pre-processing into one stage with consistent time frames for all features, and employs a simple backend that directly maps greater deviations between test and reference signals to lower quality scores.

\subsubsection{Binaural extensions of PEAQ}
\label{sec:BinPEAQ}
For this investigation we also extend PEAQ by integrating two distinct binaural models that naturally expand its basic and advanced operational modes. For the basic mode we extend the DFT-analysis based timbre quality assessment system with the DFT-based binaural cue extraction model from Kämpf et al. \cite{kmpf2010standardization}, which includes distortion metrics for ITD, interaural cross-correlation (IACC), and ILD. For the advanced operational mode extension, we extend the timbre quality assessment system PEAQ-CSM, which makes use of PEAQ's advanced MOVs, primarily based on filterbank analysis, by integrating the filterbank-based binaural cue extraction model proposed by Seo et al. \cite{seo2013perceptual}. This model was chosen due to its strong correlation with subjective quality scores across various listening test databases \cite{delgado2023design}.

\subsection{Combination of binaural and monaural models into a quality score}

All studied metrics combine monaural (and possibly binaural) features to generate a unified quality score. For example, the developers of MoBi-Q found that the overall audio quality is primarily influenced by the lower quality aspect in a signal, whether monaural or binaural. Consequently, they proposed a method to derive the overall quality by selecting the output exhibiting the greatest quality degradation, represented mathematically as: 
\begin{multline}
\label{eq:MoBiCombination}
\small
\mathrm{MoBiQ}
  = \min\Bigl(\log_{10}(0.0528 \times \mathrm{OPMdual}),\\
              0.0078 \times \mathrm{binQ}\Bigr)
\end{multline}
This estimate was validated by comparison to subjective quality scores from a listening test consisting of 16 normal hearing listeners evaluated a total of 119 audio items, which included various combinations of monaural and binaural distortions. Similar to MoBi-Q, e-MoBi-Q also selects the lower-quality component of the available monaural and binaural features as the overall quality rating, based on the results of MoBi-Q.

In the case of the binaural extensions of PEAQ, we combine the output quality scores and the binaural cues described in \Section{sec:BinPEAQ} using Multivariate Adaptive Regression Splines (MARS), as implemented in \cite{Jekabsons_areslab}, to form an overall timbre and spatial audio quality score using the listening test signals and subjective scores of the Unified Speech and Audio Coding (USAC) Verification Tests (VT) 2 and 3 \cite{USACdatabase} as training data. These databases feature stereo-format reference signals, comprising 24 items each, divided equally into speech, music, and mixed content. The tests evaluated various coding technologies, including USAC, AMR-WB\mbox{+}, and HE-AACv2. USAC VT2 specifically focuses on stereo signals with intermediate to high impairments in the stereo field, resulting from the use of parametric stereo tools by low bit-rate codecs operating between 16 and 24 kbps. In contrast, USAC VT3 contains stereo signals with smaller impairments, achieved through higher bitrates ranging from 32 to 96 kbps, where codecs do not typically rely on parametric models.

The following section compares the performance of the mentioned objective quality assessment systems based on monaural and binaural models. For the quality metrics based on binaural models, we also report the performance results of their monaural/timbre assessment modules (available as independent outputs) separately to show how much they contribute to overall performance in estimating basic audio quality.

\section{Experiment}

\subsection{Listening test description}
\label{sec:lt_experiment}

This section describes briefly the listening experiments carried out for the extension of ODAQ documented in detail in  \cite{dick2025ODAQ}, a MUlti Stimulus test with Hidden Reference and Anchor (MUSHRA) listening test was conducted at Ball State University in the US with 16 undergraduate student participants (mean age 20.8 years) from Telecommunications or Music Media Production programmes, utilising Beyerdynamic DT770 Pro 250 Ohm headphones in acoustically damped rooms.

\begin{figure}[t]
\centering
\resizebox{0.49\textwidth}{!}{%
\includegraphics{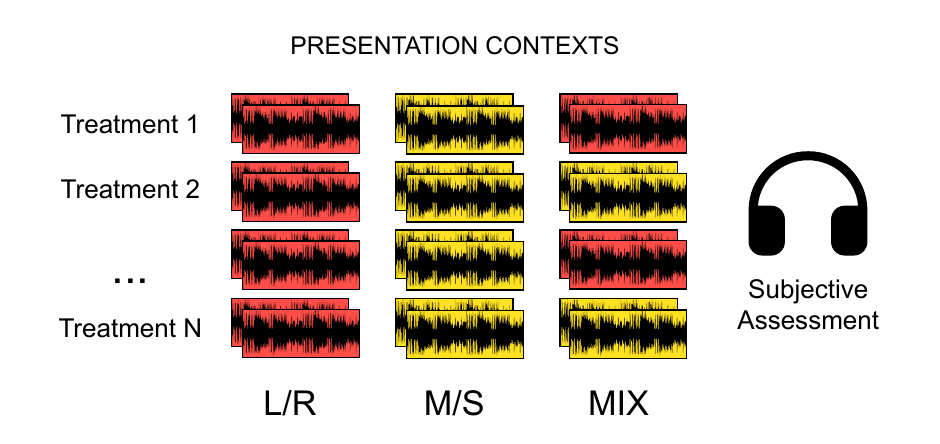}
}
\caption{Experimental conditions for the degraded files. In the MUSHRA test, listeners were presented with LR-only degradations (applied independently to the left and right channels), MS-only degradations (applied to the mid and side channels), or a mixture of both.}
\label{fig:Presentations}
\end{figure}

The used test signals were 48kHz/24-bit stereo excerpts 10 to 16 seconds long and varied in stereo characteristics. Table~\ref{tab:ItemDescription} describes the content of the excerpts, which includes solo instruments (\texttt{violin}, \texttt{glock}) centered in the image, wider stereo music mixes (\texttt{Pop}, \texttt{RnB}), and artificially mixed dialog and music items with hard-panned dialog in one of the channels (\texttt{panDialogM}, \texttt{panDialogF}). Two monaural coding artifacts, Spectral Holes (SH) (simulated by randomly inserted spectral holes with a modified perceptual audio codec) and Quantization Noise (QN) (simulated by additive noise with a frequency-dependent noise-to-mask ratio), were adapted from the original ODAQ study to also generate distortions in the stereo image. To achieve this, these distortions were applied as treatments to the audio excerpts at five quality levels (Q1-Q5) \cite{Torcoli2024ODAQ} using LR and MS stereo preprocessing. The experiment included trials for the exclusive comparison of LR-only or MS-only conditions. In addition, mixed trials (SHmix, QNmix) were also included where both LR and MS versions (two quality levels each) were presented for direct comparison to assess presentation context (see \Figure{fig:Presentations}). All 22 trials, each containing seven test conditions and a hidden reference, included MUSHRA standard 3.5 kHz and 7.0 kHz lowpass anchors, with an additional mono anchor in the SHmix and QNmix trials to the evaluate spatial image contribution of the conditions.

\begin{table}[t]
\centering
\begin{tabular}{|p{3.5cm}|p{3cm}|}
\hline
\textbf{Excerpt Label} & \textbf{Description}\\
\hline
\texttt{Violin} & Solo violin \\
\hline
\texttt{Glock} & Solo glockenspiel \\
\hline
\texttt{RnB} & RnB music \\
\hline
\texttt{Pop} & Instrumental pop music \\
\hline
\texttt{panDialogF} & Hard-panned female speech and background music\\
\hline
\texttt{panDialogM} & Hard-panned male dialog with background music\\
\hline
\end{tabular}
\caption{Labels and description of the audio excerpts used in the listening experiments described in \Section{sec:lt_experiment}.}
\label{tab:ItemDescription}
\end{table}

\subsection{Pre-processing of audio files for objective quality assessment}

The listening test files are already time-aligned, and they underwent individual level adjustments based on assumed listening levels by each metric. Specifically, PEAQ and its derivatives assume a presentation level of 92 dB SPL, while HAAQI, MoBi-Q and eMoBI-Q assume 65 dB SPL. For PEMO-Q, a signal amplitude of 1 is considered to correspond to 100 dB SPL. Furthermore, all models accept files at 48 kHz, except for HAAQI, which automatically resamples signals to 24 kHz in normal hearing mode.

\subsection{Objective quality assessment system validation}

The respective quality estimates of the studied objective systems mentioned in \Section{sec:metrics} were evaluated in terms of Pearson correlation with their subjective score counterparts. Low-pass anchors present in the listening test are excluded from the analysis. As the objective metrics estimates were calibrated to different quality scales, a third order mapping was applied prior to the correlation calculation to compensate for possible biases and offsets stemming from each of the system's scales and the listening test scores, as recommended in \cite{EvalObjective}. Furthermore, we investigate how the predictive power of various objective metrics depends on the input audio excerpt, taking into account the different treatments (QNMS, QNLR, SHMS, SHLR) and their presentation context. The reported pooled mean correlation values were calculated in Fisher's z-domain.

\section{System validation results}

In the following Section, we report the studied systems prediction performance on listener ratings for each experiment. We analyze the proposed factors of QN and SH artifacts in LR and MS configurations, along with the presentation context used in the subjective quality tests. We also break down results by audio excerpt, distinguishing two relevant categories: quality prediction on signals without hard-panned objects, and prediction on signals with hard-panned objects (for example, speech on one channel while background music plays on both channels). The motivation behind this categorization is that, when an auditory object appears in only one stereo channel, the similarity between left and right signals is significantly reduced. This phenomenon poses a challenge for objective metrics that assume both channels carry similar content by computing a single quality score that results from averaged distortion metrics across left and right channels.

\subsection{Results per experiment: artifact type and presentation context}

\begin{figure}[t]
\centering
\resizebox{0.49\textwidth}{!}{%
\includegraphics{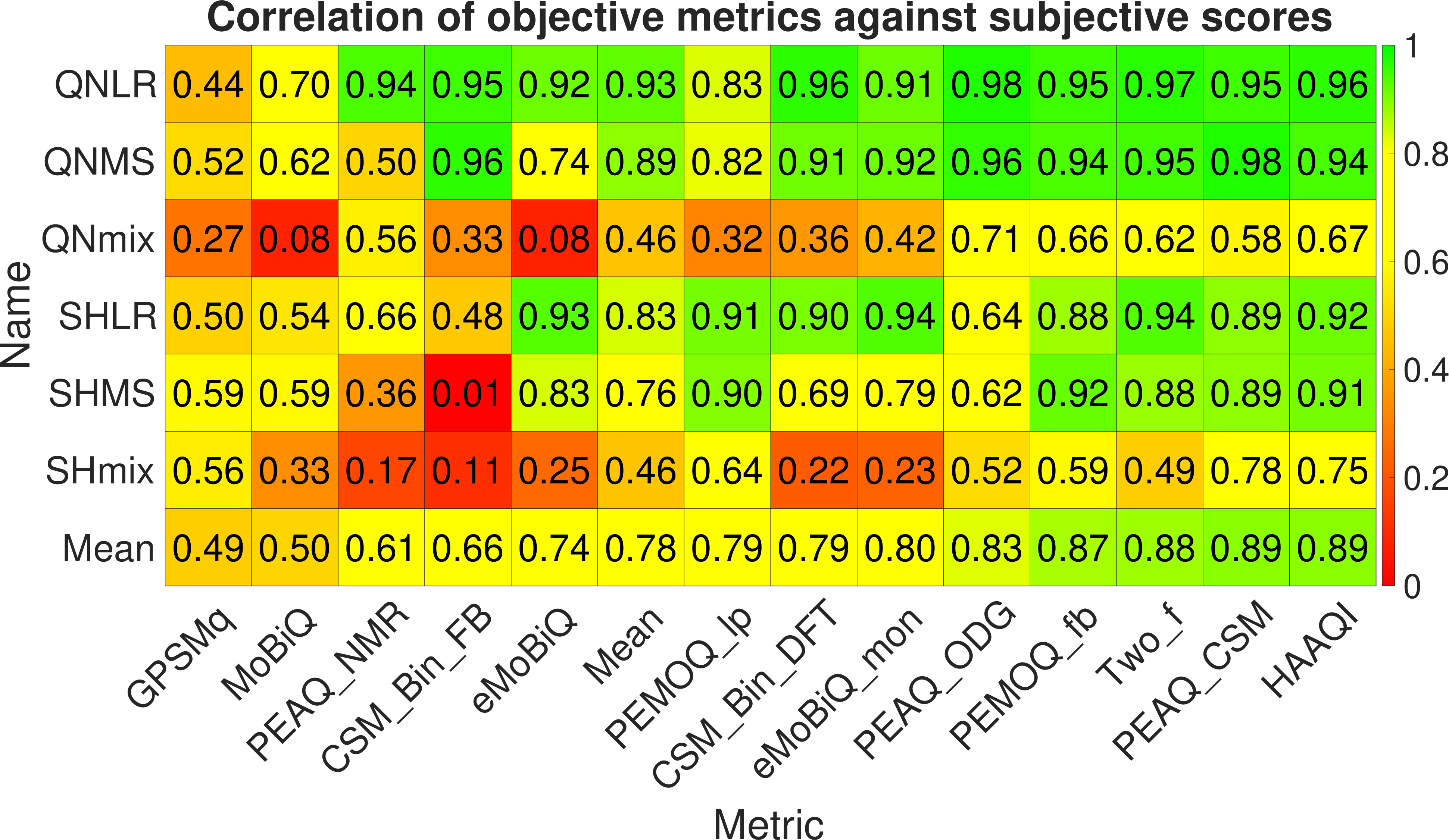}
}
\caption{Correlation between objective metric predictions and subjective quality scores for each experiment, taking into account artifact type and presentation context. $\mbox{CI}_{95\%} \leq \pm  0.01$ for all estimates.}
\label{fig:Results_per_experiment}
\end{figure}

Our analysis of the average correlation values (\Figure{fig:Results_per_experiment}) from the experiments reveals clear differences in the prediction performance of objective quality assessment models. Interestingly, the models demonstrating the best overall performance in this context do not explicitly incorporate dedicated spatial audio distortion metrics or comprehensive binaural models. Conversely, objective measures that include binaural distortion metrics for spatial audio tend to exhibit below-average performance for the most part.

Also, it can be observed that the performance of objective quality metrics is generally low for the experiments SHmix and QNmix, with many correlation values against subjective experiments being weak (below 0.8), and the average correlations for SHmix and QNmix (0.44) being notably low.  This is in contrast to the overall strong performance (with many correlations well above 0.8, and frequently exceeding 0.9) observed for various objective metrics when assessing the same distortions presented in homogeneous contexts (QNLR, QNMS, SHLR, and SHMS).

\subsection{Results per audio excerpt}

\begin{figure}[t]
\centering
\resizebox{0.49\textwidth}{!}{%
\includegraphics{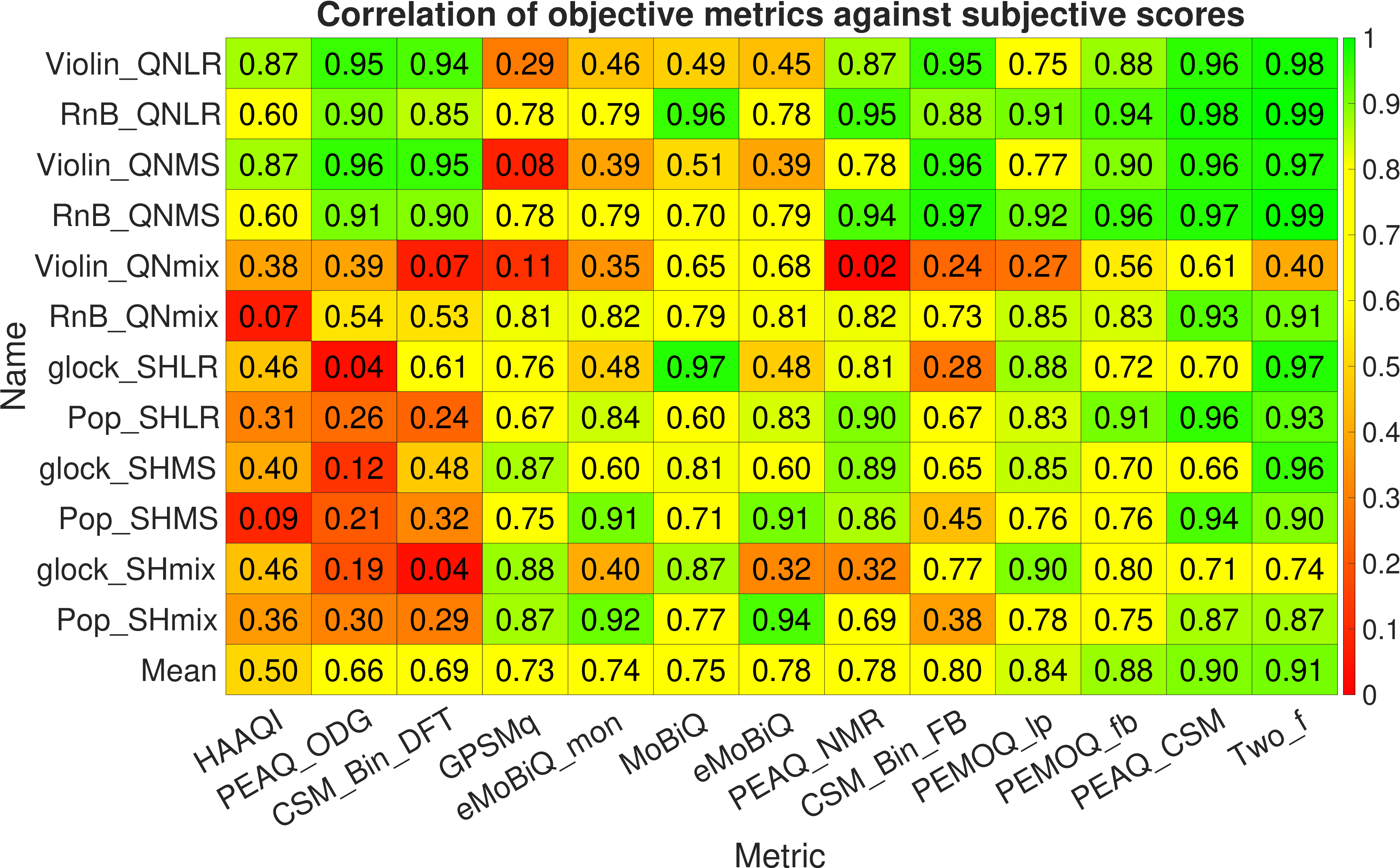}
}
\caption{Correlation between objective metric predictions and subjective quality scores for each audio excerpt and their respective treatments. Audio excerpts without hard-panned auditory objects. $\mbox{CI}_{95\%} \leq \pm 0.01$ for all estimates.}
\label{fig:NO_HARDPAN_ITEMS}
\end{figure}

\begin{figure}[t]
\centering
\resizebox{0.49\textwidth}{!}{%
\includegraphics{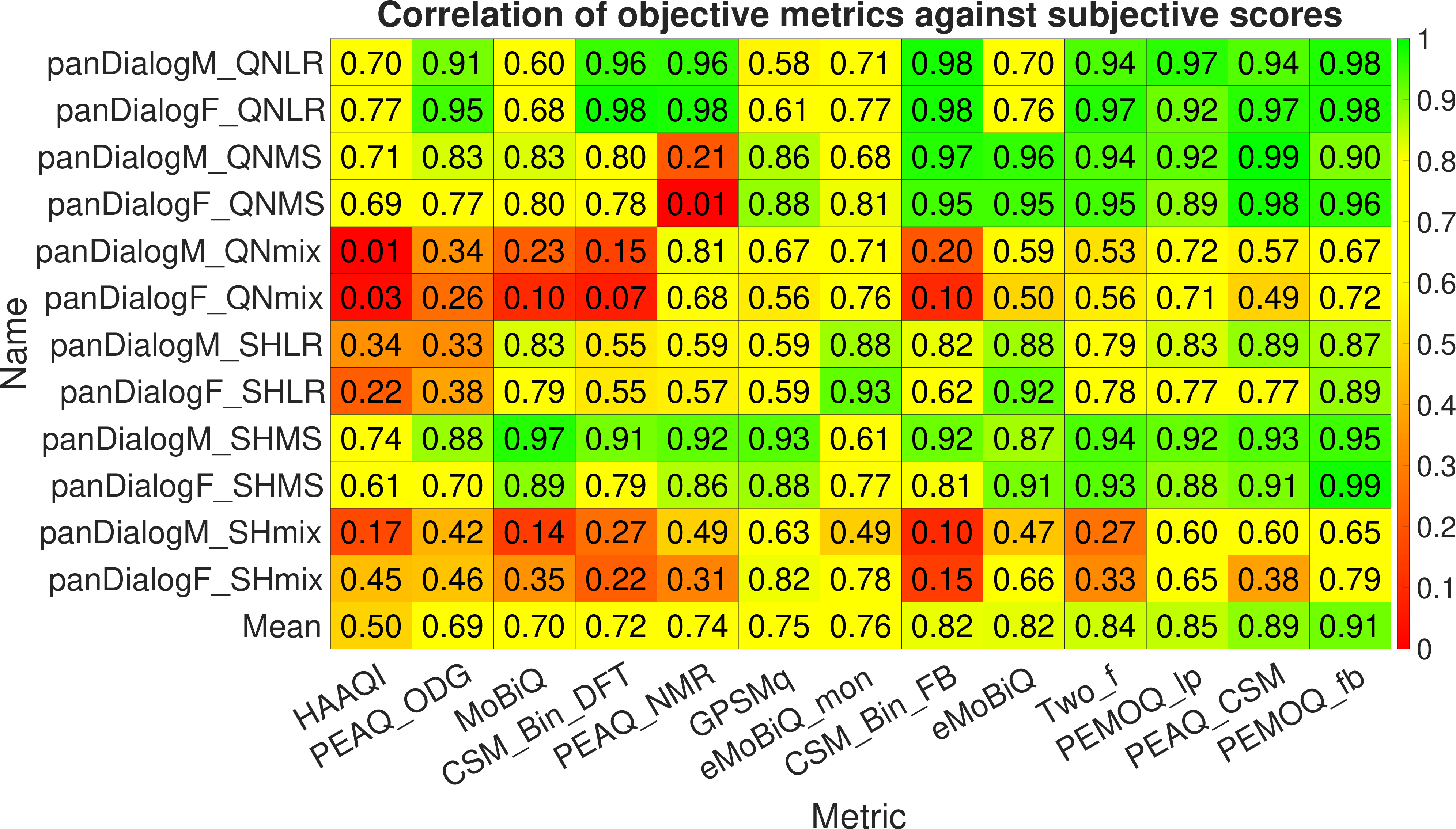}
}
\caption{Correlation between objective metric predictions and subjective quality scores for each audio excerpt and their respective treatments. Audio excerpts with hard-panned auditory objects. $\mbox{CI}_{95\%} \leq \pm 0.01$ for all estimates.}
\label{fig:HARDPAN_ITEMS}
\end{figure}

Figures~\ref{fig:NO_HARDPAN_ITEMS} and~\ref{fig:HARDPAN_ITEMS} show the system prediction results per audio excerpt, considering non-hard-panned items and hard-panned items respectively. 

Figures~\ref{fig:NO_HARDPAN_ITEMS} and~\ref{fig:HARDPAN_ITEMS} illustrate the system prediction results per audio excerpt, specifically distinguishing between non-hard-panned and hard-panned items, respectively. An analysis of these results reveals that certain metrics, such as PEAQ-CSM and PEMOQ, demonstrated equally acceptable mean performance for both hard-panned and non-hard-panned items. However, all objective metrics generally struggled in predicting quality for almost all SHmix and QNMix items, with only a few exceptions. For metrics like HAAQI and PEAQ, their performance was notably poor, despite having yielded very good results in overall quality assessment of the different experiments (as shown in \Figure{fig:Results_per_experiment}). This disparity indicates that, while these metrics can predict the general trends in overall quality scores across the entire database of distorted audio files, they struggle to accurately rank the degradations of individual items when examined in isolation. This underscores the critical importance of conducting both item-wise and experiment-wise analyses. The Two-F model showed better performance for non-hard-panned items, experiencing a loss of performance when dealing with hard-panned items, particularly in mixed conditions (QN, SH) and SHLR.

When considering the quality metrics that use binaural models, a general observation was that they did not significantly enhance performance beyond the timbre baseline, and sometimes even worsened performance. This trend is evident when comparing GPSMq against MoBIQ, eMoBiQ\_mon against eMoBiQ, and PEAQ-CSM against CSM\_Bin\_FB.

Finally, the NMR metric, which had previously demonstrated the best performance on previous studies \cite{Torcoli2024ODAQ}, experienced a catastrophic failure in predicting hard-panned dialog quality on QNMS, in addition to the general limitations shared by all the metrics for the SHmix and QNmix experiments. Despite this, the same hard-panned dialog item was predicted without problems under QNLR conditions.

\section{Discussion}
\subsection{Best-performing metrics}
The results showing the strong performance of Two-F, PEMO-Q (filterbank), and PEAQ-CSM—all of which depend heavily on modulation-distortion metrics—underscore that accurate modeling of modulation distortion is essential for predicting these degradations. Other systems that did not perform as well, such as basic PEAQ, also include modulation-distortion models but apparently assign them less weight in their overall quality score.

\subsection{Limitations of binaural metrics and their combination with timbre quality models}
The results also showed that, despite leveraging spatial cue distortion metrics, binaural models rarely outperform monaural baselines in predicting basic audio quality. This reflects a well‐documented listener bias: quality judgments tend to hinge far more on timbral fidelity (distortion, spectral balance) than on spatial attributes \cite{rumsey2005relative}. Empirical studies have reported that timbral fidelity contributes roughly twice as much as spatial fidelity to overall quality ratings. This bias may be even stronger for the particular types of distortions present in this database. Consequently, even if a binaural model perfectly detects spatial distortions, it may yield little improvement in prediction accuracy, since listeners generally do not perceive spatial degradations as major quality impairments. 

Another reason for the limited performance may be due to the way binaural models and monaural models are combined. None of the different combination methods mentioned in \Section{sec:HandlingOfStereoSignals}, including heuristic methods (e.g., Equation~\ref{eq:MoBiCombination}) or machine-learning-based methods (for the PEAQ extensions) seem to have brought an increase of performance on the resulting models. These results suggest that the combination of monaural and binaural distortion metrics into a meaningful quality prediction score still remains a challenge.

\begin{figure}[t]
\centering
\resizebox{0.49\textwidth}{!}{%
\includegraphics{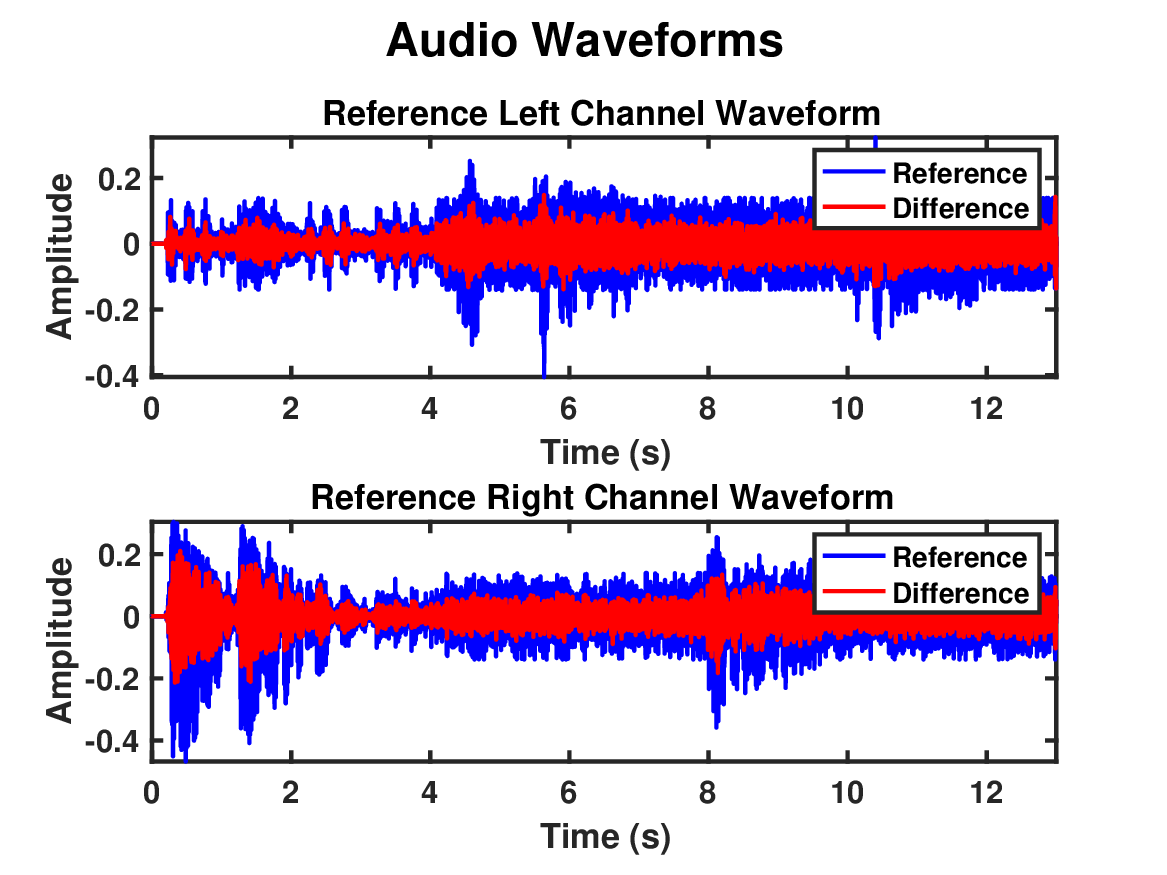}
}
\caption{Reference and difference (to degraded signal) signal for the item \texttt{panDialogF} and a QNLR degradation maintaining a NMR of 6dB.}
\label{fig:PLOT_DIFFERENCE_LR}
\end{figure}
\begin{figure}[t]
\centering
\resizebox{0.49\textwidth}{!}{%
\includegraphics{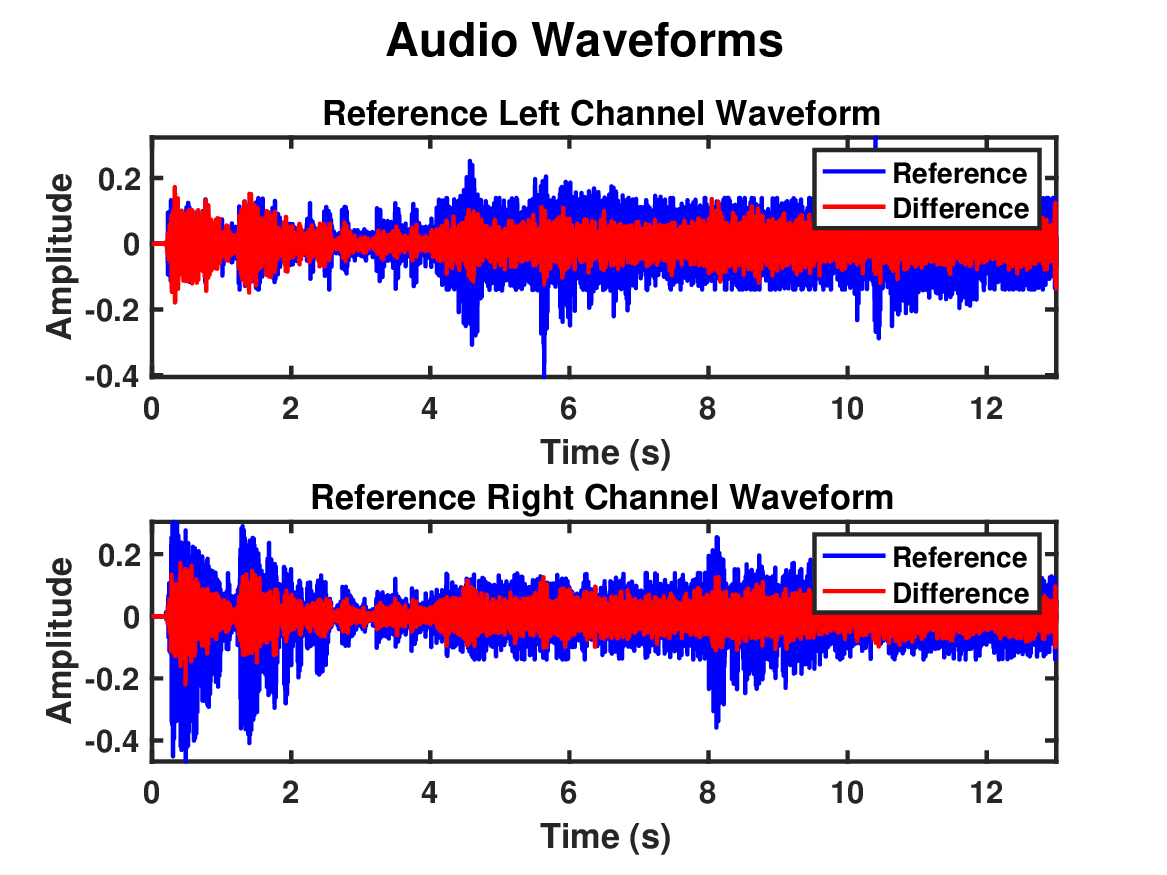}
}
\caption{Reference and difference (to degraded signal) signal for the item \texttt{panDialogF} and a QNMS degradation maintaining a NMR of 6dB.}
\label{fig:PLOT_DIFFERENCE_MS}
\end{figure}

\subsubsection{Failure of NMR on hard-panned items with MS quantization noise}
A likely reason for the NMR metric failure on the hard-panned items with QNMS, which is purely based on the concept of auditory masking of noise, seems to lie in how the quantization noise (QN) distortion interacts with the different stereo processing configurations. In the QNLR configuration, QN is introduced independently into each channel, resulting in a noise signal that closely resembles the original signal in those channels (see \Figure{fig:PLOT_DIFFERENCE_LR}), thus allowing the NMR metric to perform effectively.

However, when QN is added to the Mid signal in the QNMS configuration, as illustrated in \Figure{fig:PLOT_DIFFERENCE_MS}, the noise signal becomes similar to the Mid original signal. For hard-panned items, the left and right signals are distinctly different from the Mid signal, and consequently, also differ significantly from this noise signal. This dissimilarity complicates the NMR calculation, causing the metric to overestimate the perceived disturbance of the noise beyond the masking thresholds, which are individually calculated for the left and right channels. As a result, the averaged NMR across the left and right channels does not correlate with the quality degradation reported by listeners. This illustrates a general challenge for metrics that derive a single quality score by averaging distortion metrics across left and right channels, especially when dealing with hard-panned auditory objects where channel similarity is significantly reduced. Metrics that do not exclusively derive their quality degradation estimates from masking principles seem to be more robust when facing these configurations. 

\subsubsection{Influence of presentation context}
Finally, the results showed that almost all of the metrics struggle with quality prediction on the QNmix and SHmix experiments, but some of the quality metrics reliably predict the same signals and distortions when presented in an homogeneous context. This discrepancy underscores that objective metrics must always be validated against subjective scores to accurately evaluate their prediction performance. Even though listening tests are considered the gold standard, their design plays a crucial role. The results suggest that in the experiment's specific design, listeners might assign different quality scores to identical signals depending on the presentation context. The presentation context of the same signals and distortions appears to confound almost all objective metrics, hinting to the fact that these metrics fail to implicitly or explicitly model this factor in their analysis models. Future developments in audio quality metrics need to address this limitation. Although it is not clear how presentation context can be explicitly modeled into the metrics (a top-down process), a potential solution may include a data-driven approach that incorporates ground truth sets with different presentation contexts to map the different distortion metrics (i.e., bottom-up processes) into a quality score estimate. 

\section{Conclusion}

Our observations from this investigation point toward an interesting discovery: contrary to the common assumption that accurately predicting subjective quality for stereo processing invariably requires sophisticated spatial metrics, certain timbre quality metrics appear to yield satisfactory predictions in many instances. However, two scenarios challenge this premise: highly context-dependent listening tests, and, for some metrics, cases where the artifacts in the stereo channels are introduced in items with strong left-right imbalance in terms of auditory events (e.g., hard-panned items). By recognizing and isolating these special cases, future research can refine models to handle the interplay between timbral and spatial elements more effectively.

\section*{Acknowledgements}

We would like to sincerely thank Mhd Modar Halimeh for the valuable discussions on objective metrics and listening test design for this work.

\bibliographystyle{IEEEtran}
\begin{small}
\bibliography{refs}
\end{small}

\end{document}